\documentstyle[preprint,aps,prd]{revtex}
\tightenlines

\newcommand{\be}{\begin{equation}}
\newcommand{\ee}{\end{equation}}
\newcommand{\bn}{\begin{eqnarray}}
\newcommand{\en}{\end{eqnarray}}
\newcommand{\bd}{\begin{displaymath}}
\newcommand{\ed}{\end{displaymath}}

\begin{document}
\draft

\title{\bf Thermal fluctuations of a quantized massive
scalar field in Rindler background}

\author{ R. Medina \footnote{e-mail: rmedina@cpd.efei.br}
and\hspace{0.25cm} 
E. S. Moreira Jr. \footnote{e-mail: moreira@cpd.efei.br}}
\address{Instituto de Ci\^{e}ncias\\
         Escola Federal de Engenharia de Itajub\'{a}\\
         Av. BPS 1303, 37500-903, Itajub\'{a}, Minas Gerais, Brazil}
\date{November, 2000}
\preprint{hep-th/00?????}

\maketitle
\begin{abstract}
Thermal fluctuations
for a  massive scalar field in the Rindler wedge
are obtained by applying the point-splitting procedure
to the zero temperature Feynman propagator in a
conical spacetime. Renormalization is implemented by removing
the zero temperature contribution. It is shown that
for a field of non vanishing mass the thermal fluctuations,
when expressed in terms
of the local temperature, do not have Minkowski form.
As a by product, Minkowski vacuum fluctuations seen by an uniformly
accelerated observer are determined and confronted with
the literature.

\end{abstract}

\narrowtext

\section{Introduction}
\label{int}
It is well known that certain aspects of
the study of
quantum fields vibrating near
the horizon of a large black hole
amount to consider thermal averages in the
Rindler wedge \cite{dav82,wal84}.
Such a connection is one of the reasons which makes the
latter a topic of much interest \cite{tho85,sus94}.

In this letter,
the thermal fluctuations
$\langle\phi ^{2}(x)\rangle$ for a massive scalar
field in the Rindler wedge are worked out by
applying an algorithm which consists of reading Rindler
thermal averages from vacuum averages in a
conical spacetime \cite{dow77,dow78,dow94}.
In Sec. \ref{min}, the thermal fluctuations
in Minkowski spacetime are presented in order
to be compared  with those in the Rindler wedge.
Sec. \ref{alg} explains the algorithm mentioned above.
In  Sec. \ref{rin}, 
$\langle\phi ^{2}(x)\rangle$ in four dimensional
Rindler background are determined in various temperature
regimes, and for small and large masses of the field.
As quantities in Sec. \ref{rin} are renormalized by
subtracting their zero temperature values (the Rindler vacuum
contribution), for a particular temperature (the Hawking temperature)
they should reproduce Minkowski vacuum fluctuations as seen
from an uniformly accelerated frame. This fact is verified in
Sec. \ref{acc}. Final remarks are presented in Sec \ref{dis},
and an appendix, where the Euler-Maclaurin formula is used to
perform a summation, closes the letter.

Throughout the text, units are such that  $\hbar=c=k_{B}=1$.

\section{Thermal $<\phi^{2}>$ in Minkowski spacetime}
\label{min}
In Minkowski spacetime,
the Feynman propagator for a scalar field $\phi(x)$, of mass $M$,
in thermal equilibrium with a heat bath at temperature $T$ is
given by (see e.g. Ref. \cite{lan98})
\be
G_{{\cal F}}(x,x')=
G_{{\cal F}}^{0}(x,x')
-i\int_{-\infty}^{\infty}
\frac{d^{4}{\rm k}}{(2\pi)^{3}}\
\exp\{-i{\rm k}({\rm x}-{\rm x}')\}\
\frac{\delta({\rm k}^{2}-M^{2})}{e^{|{\rm k_{0}}|/T}-1},
\label{minp}
\ee
where $G_{{\cal F}}^{0}(x,x')$ is the familiar zero temperature
Feynman propagator.
According to the point-splitting procedure \cite{dav82},
the average $<\phi^{2}>$ can be obtained from the Feynman
propagator $G_{{\cal F}}(x,x')$ by considering
\be
\langle\phi ^{2}(x)\rangle=
i\lim_{x'\rightarrow x}G_{{\cal F}}(x,x'),
\label{vev2}
\ee
where renormalization is implemented by dropping
the zero temperature contribution. Thus one considers
Eq. (\ref{minp}) in Eq. (\ref{vev2}), and after some
manipulations it results
\be
\langle\phi ^{2}(x)\rangle=
\frac{1}{2\pi^{2}}\int_{0}^{\infty}
\frac{\left(\omega^{2}+2M\omega\right)^{1/2}}
{e^{M/T}e^{\omega/T}-1}\ d\omega.
\label{vev1}
\ee
For high temperatures ($M/T\ll 1$)
and low temperatures  ($M/T\gg 1$)     ,
$\langle\phi ^{2}(x)\rangle$ 
behaves as ($\gamma$ is the Euler constant)
\be
\frac{T^{2}}{12}-\frac{MT}{4\pi}
-\frac{M^{2}}{8\pi^{2}}
\left(\log \frac{M}{4\pi T}+\gamma-\frac{1}{2}\right)
\label{mht}
\ee
and
\be
M^{1/2}\left(\frac{T}{2\pi}\right)^{3/2}
\exp \left\{-\frac{M}{T}\right\},
\label{mlt}
\ee
respectively.

\section{Thermal averages in Rindler background}
\label{alg}
In order to obtain thermal effects at temperature $1/2\pi\alpha$ in
the quantum mechanics of fields in
Rindler background (the right Rindler wedge, for which $\rho$ below is
positive), 
\be
ds^{2}=\rho^{2} d\eta^{2}-d\rho^{2}-dx^{2}-dz^{2},
\label{rindler}
\ee
one analytically
continues the time coordinate $\eta$ to imaginary values,
taking $\varphi:=i\eta$ periodic with period $2\pi\alpha$.
Considering further $t:=ix$,
Eq. (\ref{rindler}) 
leads to the
geometry of a four dimensional conical spacetime.
Namely, the line element is recast as
$ds^{2}=dt^{2}-d\rho^{2}
-\rho^{2}d\varphi^{2}-dz^{2} $
and the identification
$(t,\rho ,\varphi,z)
\sim  (t,\rho ,\varphi +2\pi\alpha ,z)$
holds.
Hence it seems that one can read from vacuum averages
in conical spacetime thermal averages in the Rindler wedge.
Indeed this is the case, as has been shown in Refs. \cite{dow77,dow78,dow94}.
It happens that the zero temperature propagator
$G_{{\cal F}}^{(\alpha)}(x,x')$
computed in a conical spacetime
may be interpreted as Rindler thermal propagator
for temperature $1/2\pi\alpha$.
It follows that since
$G_{{\cal F}}^{(\alpha=1)}(x,x')$
is the usual propagator in Minkowski spacetime,
this  propagator is a
Rindler thermal propagator for temperature $1/2\pi$.
Therefore the Minkowski vacuum, which is a pure state,
is also a Rindler thermal state
(more precisely a statistical mixture),
a fact fairly well known.

\section{Thermal $<\phi^{2}>$ in Rindler background}
\label{rin}

Usually quantities in a conical spacetime are
renormalized with respect to the Minkowski vacuum
($\alpha=1$ contribution).
In the present context, it seems more natural
to renormalize them with respect
to the Rindler vacuum (zero temperature contribution,
$\alpha\rightarrow\infty$)
instead \cite{dow78,dow94}
(perhaps the best motivation for doing so
resides in the fact that Rindler thermal potentials
coincide with those obtained by a counting of states 
procedure \cite{sus94}).
One
implements that simply by subtracting from
$G_{{\cal F}}^{(\alpha)}(x,x')$ its
value for $\alpha\rightarrow\infty$, before
using prescription (\ref{vev2}), resulting
\be
\langle\phi ^{2}(x)\rangle=
i\lim_{x' \rightarrow x}
\left[
G_{{\cal F}}^{(\alpha)}(x,x')-
G_{{\cal F}}^{(\alpha=\infty)}(x,x')
\right].
\label{vev3}
\ee

\subsection*{Small masses}

It has been shown in Ref. \cite{mor95} that the Feynman propagator
for a massive scalar field in an $N$-dimensional conical
spacetime is given by
\bn
&&G_{{\cal F}}^{(\alpha)}(\Delta)=
\frac{1}{i\alpha(2\pi^{1/2})^{N}\rho^{N-2}}
\nonumber
\\&&
\quad\times\sum_{m=-\infty}^{\infty}
e^{im\Delta/\alpha}
\left\{\frac{\Gamma\left[(N-2)/2+|m|/\alpha\right]
\Gamma\left[(3-N)/2\right]}
{\pi^{1/2}\Gamma\left[(4-N)/2+|m|/\alpha\right]}
\nonumber\right.
\\
&&\quad\qquad\qquad\times\left.
\ {}_{1}F _{2}\left[\frac{3-N}{2};\frac{4-N}{2}-
\frac{|m|}{\alpha} ,
\frac{4-N}{2}+\frac{|m|}{\alpha} ;(M\rho)^{2}\right]\nonumber\right.
\\
&&\quad\qquad\qquad\left.+\;2^{-2|m|/\alpha}
(M\rho)^{N-2(1-|m|/\alpha)}
\frac{\Gamma\left[(2-N)/2-|m|/\alpha\right]}
{\Gamma\left[1+|m|/\alpha\right]}\nonumber\right.
\\
&&\left.\quad\qquad\qquad\times
\ \ {}_{1}F_{2}\left[\frac{1}{2}+\frac{|m|}{\alpha};
\frac{|m|}{\alpha}+\frac{N}{2} ,
1+\frac{2|m|}{\alpha} ;(M\rho)^{2}\right]
\right\},
\label{del}
\en
where ${}_{1}F_{2}\left[a ;b,c;z\right]$
denotes the generalized hypergeometric function,
$\Delta:=\varphi-\varphi'$
and one has set
$t=t'$,  $\rho=\rho'$ and $z=z'$.
In a four-dimensional background ($N=4$),
when $\alpha$ is not very large and $M\rho\ll 1$,
in order to obtain mass corrections
of order $(M\rho)^{2}$, one has to consider
in the second term inside the curly brackets of Eq. (\ref{del})
only the contribution corresponding to $m=0$.
Then dimensional regularization leads to \cite{mor95}
(terms which vanish in the limit $\Delta\rightarrow 0$
will be omitted)
\be
G_{{\cal F}}^{(\alpha)}(\Delta)= {\rm u.v.}
-\frac{i}{48\ \pi^{2}\alpha^{2}\rho^{2}}
-\frac{iM^{2}}{8\pi^{2}}
\left[\frac{1}{\alpha}\left(\log \frac{M\rho}{2}
+\gamma-\frac{1}{2}\right)-\log\alpha\right],
\label{newp}
\ee
where
\be
{\rm u.v.}:=
-\frac{i}{4\pi^{2}\rho^{2}\Delta^{2}}
-\frac{iM^{2}}{8\pi^{2}}\log \Delta
\label{uv}
\ee
gives rise to the ultraviolet divergences when the
limit in Eq. (\ref{vev3}) ($\Delta\rightarrow 0$) is
taken into account.
Using Eq. (\ref{newp}) in  Eq. (\ref{vev3}) when
$M=0$ one reproduces the known expression,
$<\phi^{2}>=1/48\ \pi^{2}\alpha^{2}\rho^{2}$. 
Taking into account that the local (Tolman)
temperature 
${\cal T}:=1/\beta \sqrt{g_{00}}$ 
(where $1/\beta$ is the 
heat bath temperature \cite{lan63})
in the Rindler wedge is given by 
\be
{\cal T}:=\frac{1}{2\pi\alpha\ \rho},
\label{ttem}
\ee
the massless $<\phi^{2}>$ can be recast as ${\cal T}^{2}/12$,
and hence is Minkowski (cf. Eq. (\ref{mht})).
For the massive case Eq. (\ref{vev3})
seems to lead to an inconsistency due to the presence
of the term containing $\log\alpha$ in Eq. (\ref{newp}) which
diverges in the limit $\alpha\rightarrow\infty$.
The following analysis shows that when
$\alpha\rightarrow\infty$
new terms of order $(M\rho)^{2}$ arise in
Eq. (\ref{del}), compensating the logarithmic divergence and
yielding a finite result.

In the limit of very large $\alpha$ not only the $m=0$ contribution
that arises from the
second term inside the curly brackets of Eq. (\ref{del})
is relevant to compute mass corrections of order $(M\rho)^{2}$.
One must also consider (for $N=4$)
\be
\frac{1}{i\alpha (4\pi)^{2}\rho^{2}}\times E(\Delta),
\label{correction}
\ee
where
\bn
E(\Delta)&:=&\sum_{m\neq 0}e^{im\Delta/\alpha}
2^{-2|m|/\alpha}
(M\rho)^{2(1+|m|/\alpha)}
\frac{\Gamma\left[-1-|m|/\alpha\right]}
{\Gamma\left[1+|m|/\alpha\right]}
\nonumber\\
&=&2(M\rho)^{2}\sum_{m=1}^{\infty}a_{m}\cos\frac{m\Delta}{\alpha},
\label{edelta}
\en
with
\be
a_{m}:=\frac{\pi\csc\{m\pi/\alpha\}}
{(1+m/\alpha)\Gamma^{2}[1+m/\alpha]}
\left(\frac{M\rho}{2}\right)^{2m/\alpha}.
\label{am}
\ee
In order to determine $E(\Delta)$ for large values
of $\alpha$, one applies
the method outlined in the Appendix.
Then adding Eq. (\ref{correction}) to
Eq. (\ref{newp}) it follows (only terms
up to $1/\alpha^{2}$ order will be kept)
\bn
&&G_{{\cal F}}^{(\alpha)}(\Delta)= {\rm u.v.}
\nonumber
\\
&&-\frac{i}{48\ \pi^{2}\alpha^{2}\rho^{2}}
+\frac{iM^{2}}{8\pi^{2}}
\left\{\frac{1}{6\alpha^{2}}\left[\left(\log \frac{M\rho}{2}
+\gamma-\frac{1}{2}\right)^{2}+\frac{1}{4}\right]
+\log\left(-\log\left(\frac{M\rho}{2}\right)^{2}
\right)\right\},
\label{newpp}
\en
which is clearly finite for $\alpha\rightarrow\infty$, yielding
\be
G_{{\cal F}}^{(\alpha=\infty)}(\Delta)= {\rm u.v.}
+\frac{iM^{2}}{8\pi^{2}}
\log\left(-\log\left(\frac{M\rho}{2}\right)^{2}
\right).
\label{newpr}
\ee

Therefore for
a massive field one has two regimes of 
heat bath temperature $1/2\pi\alpha$ to investigate.
When $\alpha$ is not very large, by subtracting Eq. (\ref{newpr})
from Eq. (\ref{newp}), and then applying prescription
(\ref{vev3}), it results
\be
\langle\phi ^{2}(x)\rangle=
\frac{{\cal T}^{2}}{12}+
\frac{M^{2}}{8\pi^{2}}
\left[2\pi\left(\log \frac{M\rho}{2}
+\gamma-\frac{1}{2}\right)\rho {\cal T}
+\log\left(-4\pi\rho {\cal T}\log\frac{M\rho}{2}\right)
\right],
\label{smalla}
\ee
where Eq. (\ref{ttem}) has been observed.
When $\alpha$ is very large, 
by subtracting Eq. (\ref{newpr})
from Eq. (\ref{newpp}), Eq.
(\ref{vev3}) leads to 
\be
\langle\phi ^{2}(x)\rangle=
\frac{{\cal T}^{2}}{12}
\left\{1-\left(M\rho\right)^{2}
\left[\left(\log \frac{M\rho}{2}
+\gamma-\frac{1}{2}\right)^{2}+\frac{1}{4}\right]\right\}.
\label{largea}
\ee
Noticing the equality 
\be
\frac{M}{{\cal T}}=M\rho\ 2\pi\alpha,
\label{ratio}
\ee
it follows that
Eq.  (\ref{smalla}) should be compared with 
Eq.  (\ref{mht}), and Eq. (\ref{largea})
should be compared with Eq. (\ref{mht}) or
Eq. (\ref{mlt}) (depending on how large $\alpha$ is). 
It is clear that even for small
masses the behaviour of the Rindler thermal fluctuations
departs from the Minkowski type.

\subsection*{Large masses}

For large masses ($M\rho\gg 1$)
Eq. (\ref{del}) is not very 
handy to evaluate thermal fluctuations.
It is rather more convenient to work 
with expression \cite{emi94}
\bn
G_{{\cal F}}^{(\alpha)}(\Delta)&=&
G_{{\cal F}}^{(\alpha=1)}(\Delta)
+\frac{iM}{16\ \pi^{3}\alpha\rho}
\int_{0}^{\infty}d\omega\ 
\frac{K_{1}\left(2M\rho\cosh\{\omega/2\}\right)}
{\cosh\{\omega/2\}}
\nonumber\\&&
\times\left[\frac{\sin\{(\Delta+\pi)/\alpha\}}
{\cosh\{\omega/\alpha\}-\cos\{(\Delta+\pi)/\alpha\}}-
\frac{\sin\{(\Delta-\pi)/\alpha\}}
{\cosh\{\omega/\alpha\}-\cos\{(\Delta-\pi)/\alpha\}}\right],
\label{emip}
\en
with $K_{1}$ denoting the modified Bessel function
of the second kind, and $\alpha>1/2$
(note that Eq. (\ref{emip}) cannot be used
to investigate the behaviour of 
$\langle\phi ^{2}(x)\rangle$ for heat bath
temperatures $1/2\pi\alpha\geq 1/\pi$).
Now, noting that for very large $\alpha$  
(in leading order)
\be
\frac{\sin\{\pi/\alpha\}}
{\alpha\left(\cosh\{\omega/\alpha\}-\cos\{\pi/\alpha\}\right)}
=
\frac{2\pi}{\pi^{2}+\omega^{2}}-\frac{\pi}{6\alpha^{2}},
\label{largeaa}
\ee
Eq. (\ref{vev3}) leads to
\be
\langle\phi ^{2}(x)\rangle=
\frac{M}{8\pi^{3}\rho}
\int_{0}^{\infty}d\omega\ 
\frac{K_{1}\left(2M\rho\cosh\{\omega/2\}\right)}
{\cosh\{\omega/2\}}
\left[
\frac{\sin\{\pi/\alpha\}}
{\alpha\left(\cos\{\pi/\alpha\}-\cosh\{\omega/\alpha\}\right)}
+
\frac{2\pi}{\pi^{2}+\omega^{2}}\right].
\label{vevla}
\ee

A closed expression for  
the Rindler thermal fluctuations at low heat bath temperatures 
$1/2\pi\alpha$ can be obtained
by inserting Eq. (\ref{largeaa}) in Eq. (\ref{vevla}).
In so doing, the integration over $\omega$ can be performed 
\cite{emi93} yielding
\be
\langle\phi ^{2}(x)\rangle=
\frac{{\cal T}^{2}}{12}
(M\rho)^{2}
\left[K_{1}^{2}(M\rho)-K_{0}^{2}(M\rho)\right],
\label{smallt}
\ee
where Eq. (\ref{ttem}) has been observed. (As a check of consistency,
by considering the behaviour of $K_{\nu}(z)$ for small $z$
in Eq. (\ref{smallt}), Eq. (\ref{largea}) is recovered.).
When $M\rho\gg 1$, one reads from Eq. (\ref{smallt}) 
a non Minkowski thermal fluctuation (cf.  Eq (\ref{mlt})) 
\be
\langle\phi ^{2}(x)\rangle=
\frac{{\cal T}^{2}}{24} \pi\exp\left\{-2M\rho\right\}.
\label{largema}
\ee

The method of steepest descent \cite{arf85}
can be used in the limit of large $M\rho$ to 
perform the integration in Eq. (\ref{vevla}).
Accordingly, considering the large $z$
behaviour of $K_{\nu}(z)$ and Eq. (\ref{ttem}), one is led to
\be
\langle\phi ^{2}(x)\rangle=
\frac{1}{8\pi^{3}\rho^{2}}
\left(1-\pi^{2}\rho{\cal T}\cot\{\pi^{2}\rho{\cal T}\}\right)
\exp\left\{-2M\rho\right\},
\label{largem}
\ee
which also departs from the Minkowski type.
(By making $\alpha$ very large in Eq. (\ref{largem}), 
Eq. (\ref{largema}) is consistently reproduced.).

\section{Accelerated frames}
\label{acc}

The world line $x(\tau)$ ($\tau$ is the proper time)
of an uniformly accelerated frame,
with proper acceleration $a$,
can be expressed in terms of Rindler coordinates
(see  Eq. ({\ref{rindler})),
where 
\be
\eta(\tau)=a\tau\hspace{2cm}\rho(\tau)=1/a
\label{acoordinates}
\ee 
($x$ and $z$ are kept constant).
The Minkowski vacuum fluctuations 
$\langle 0_{M}|\phi ^{2}(\tau)|0_{M}\rangle$
as seen from this frame can formally be 
obtained by considering the Minkowski two point function
$g(\tau-\tau'):= \langle 0_{M}|\phi(\tau)\phi(\tau')|0_{M}\rangle$
evaluated along the world line, and in the limit $\tau'\rightarrow\tau$
(i.e., $x'\rightarrow x$).
Such a procedure requires renormalization, which 
is taken as being
\be
\langle\phi ^{2}(x)\rangle=
\lim_{x'\rightarrow x}
\left[
\langle 0_{M}|\phi(x)\phi(x')|0_{M}\rangle-
\langle 0_{R}|\phi(x)\phi(x')|0_{R}\rangle
\right],
\label{aframevev}
\ee
i.e., one subtracts the Rindler vacuum contribution.
If $F(\omega)$ is the Fourier transform of $g(\tau-\tau')$,
it can be shown that \cite{tak86}
\be
\langle\phi ^{2}(x)\rangle=
\frac{1}{\pi}\int^{\infty}_{0}d\omega\ F(\omega).
\label{aframevev1}
\ee
(The Fourier transform $F(\omega)$
has been evaluated in Chapter 4
of Ref. \cite{tak86}.)

As careful manipulations 
(which will be omitted here since they are lengthy) show,
by inserting in Eq. (\ref{aframevev1}) the
expression for $F(\omega)$
in the limit of small masses 
(incidentally, it should be remarked that this limit 
is mistakenly evaluated in Ref. \cite{tak86}), one ends up with 
\be
\langle\phi ^{2}(x)\rangle=
\frac{{\cal T}^{2}}{12}+
\frac{M^{2}}{8\pi^{2}}
\left[\log \frac{M}{4\pi{\cal T}}
+\gamma-\frac{1}{2}
+\log\left(-\log\left(\frac{M}{4\pi{\cal T}}\right)^{2}\right)
\right],
\label{hsmalla}
\ee
where
\be
{\cal T}:=\frac{a}{2\pi}.
\label{htem}
\ee
Recalling Eq. (\ref{acoordinates}),
it follows that
Eq. (\ref{hsmalla}) is just
Eq. (\ref{smalla}) for $\alpha=1$.
In the limit of large masses \cite{tak86},
Eq. (\ref{aframevev1}) leads to 
\be
\langle\phi ^{2}(x)\rangle=
\frac{{\cal T}^{2}}{2\pi}\exp\left\{-\frac{M}{\pi{\cal T}}\right\}
\label{hlargema}
\ee
instead. One sees that Eq. (\ref{hlargema})
follows from  Eq. (\ref{largem})
when $\alpha=1$.

Such identifications should not be seen as
unexpected. Indeed, when $\alpha=1$,
Eq. ({\ref{vev3}) is identical to Eq. (\ref {aframevev}).

\section{Final remarks}
\label{dis}

By comparing Eq. (\ref{hsmalla}) with Eq. (\ref{mht})
and  Eq. (\ref{hlargema}) with Eq. (\ref{mlt}),
one concludes that the statement
``an accelerated observer sees the Minkowski vacuum as
a (Minkowski) thermal bath whose temperature is proportional to the 
proper acceleration'' is not correct if the scalar
field has a non vanishing mass (it should be mentioned
that this result has also been reached in Ref. \cite{tak86}
by comparing the Fourier transforms of the corresponding
two-point functions). In fact this statement
does not hold in a variety of situations 
(see Refs. \cite{tak86,jau91,chi94} and references therein).

The fact that the zero temperature propagator in a conical
spacetime can be interpreted as a Rindler thermal propagator
does not merely follow as a consequence of
recasting Eq. (\ref{rindler}) as the line element in a conical spacetime.
A crucial aspect is the requirement of finiteness of the eigenmodes at
the conical singularity, i.e. on the Rindler horizon $\rho=0$.
In fact, all the conical eigenmodes vanish at the conical
singularity, except the zero angular momentum eigenmode 
(see Ref. \cite{mor95}
and references therein). Therefore when the field $\phi(x)$
is expanded in terms
of these conical eigenmodes, the net effect is essentially the one 
corresponding to the Dirichlet boundary condition.
Rindler thermal Green functions \cite{dow77,dow78,more96},
on the other hand,
are built up with Rindler eigenmodes which, unlike the conical eigenmodes,
oscillate very rapidly as the horizon is approached. Nevertheless,
the net effect of these oscillations in the field is again 
the one corresponding to Dirichlet's boundary condition \cite{hil86}. 
(At this point it should be mentioned that Eq. (\ref{aframevev})
has been evaluated in Ref. \cite{hil86}
by using the Schr\"{o}dinger formalism. When $\alpha=1$, the first term
inside the brackets in Eq. (\ref{vevla}) drops, and the expression for
$\langle\phi ^{2}(x)\rangle$ in Ref. \cite{hil86} is reproduced.)

>From the first paragraph of Sec. \ref{rin},
one clearly sees that the Rindler thermal fluctuations
can be obtained from the corresponding
conical vacuum fluctuations (which are renormalized with respect to
the Minkowski vacuum) by subtracting from the latter their 
values for $\alpha=\infty$. When such a recipe is applied to
the expression for  
$\langle\phi ^{2}(x)\rangle$ in Refs. \cite{mor95,iel97},
a logarithmic divergence arises. As shown in Sec. \ref{rin}
above, this is just an apparent inconsistency, since the divergence
cancels out when the limit $\alpha\rightarrow\infty$ is
properly taken. By applying this recipe to the expression 
for the large mass
behaviour of the vacuum fluctuations in 
Ref. \cite{shi92}, Eq. (\ref{largem}) is reproduced
up to a factor $1/2$ (which is missing in Ref. \cite{shi92}).

A last remark concerning Eq. (\ref{largem}) 
(and also its conical counterparts in Refs. \cite{shi92,lin92})
is in order. When ${\cal T}$ approaches $1/\pi\rho$ from 
below, the right hand side of Eq. (\ref{largem})
grows without bounds. The reason for such a behaviour may be just 
a consequence of the fact that the limit of validity
of Eq. (\ref{emip}) is approached ($\alpha=1/2$), or it may be related 
with some other as yet obscured mathematical reason
(e.g., some limitation in the method of steepest descent).
This fact requires further analysis.

An investigation along these lines  
concerning the energy momentum tensor will appear
elsewhere.

The approach in this letter may be useful in the
investigation of quantum processes taking place
in massive Rindler thermal baths
(e.g. Ref.\cite{mat99}).

\acknowledgments
The authors are most grateful to Devis Iellici
for pertinent remarks, and  for useful discussions
in the early stage of this work.

\section*{APPENDIX}

One looks for an asymptotic expansion in powers 
of $1/\alpha$ for the series
\begin{eqnarray}
\frac{E(\Delta)}{2(M \rho)^2 \alpha} = \sum_{k=1}^{\infty}
H_{\alpha}(k)
\end{eqnarray}
where
\begin{eqnarray}
H_{\alpha}(x):= 
\frac{\pi}{\alpha \sin\{x \pi/\alpha\} }
\frac{ (M \rho/2)^{2x/ \alpha} \cos\{x \Delta/\alpha\}}
{\Gamma[1+x/\alpha] \Gamma[2+x/\alpha]}
\label{F}
\end{eqnarray}
The Euler-Maclaurin formula (see, for example, Chapter 7 in
Ref. \cite{Whittaker}) is appropriate for this purpose 
and makes
no use of any regularization technique. 
It may be written as 
\begin{eqnarray*}
\sum_{k=a}^{a+r} G(k) = \int_{a}^{a+r} G(x)\ dx + 
\frac{1}{2}(G(a+r)+G(a))
+ \frac{1}{12}(G'(a+r)-G'(a)) + \nonumber \\
+ \frac{1}{6} \int_{0}^{1} \phi_3 (x) 
\sum_{k=a}^{a+r-1} G'''(k+x)\ dx,
\end{eqnarray*}
where $G(x)$ is an analytic function in the interval $ [a,a+r] $, $a$ and $r$ 
are
integers ($r$ being positive) and $\phi_3 (t)$ is the third degree Bernoulli 
polynomial.
When applied to the function $H_{\alpha}(x)$ defined in Eq. (\ref{F}) for 
the
interval $(0,\infty)$, and afterwards expanding in powers of
$1/\alpha$, the Euler-Maclaurin formula leads to
\begin{eqnarray}
{\sum_{k=1}^{\infty}}^{\#} 
H_{\alpha}(k)  \sim
 \log \alpha +  C(M \rho/2, \Delta) 
- \frac{1}{\alpha} \left[\log \frac{M \rho}{2}+ \gamma -\frac{1}{2}+
{\cal O}(M\rho)\right] 
\nonumber \\
- \frac{1}{6 \alpha^2} \left[ \left(\log \frac{M \rho}{2} + \gamma
-\frac{1}{2}\right)^2 + \frac{1}{4} - \frac{\Delta^2}{4} +
{\cal O}(M\rho)\right]  \nonumber \\ + {\cal O}(1/\alpha^3) 
\cdot {\cal O} (\log^{3}(M \rho/2) ),
\label{formula}
\end{eqnarray}
where

\begin{eqnarray}
C(M \rho/2, \Delta) = 
\gamma - PV \int_{0^+}^{\infty} \log u \ \frac{d}{du} 
\left[ \frac{\pi u}{\sin\{\pi u\}} 
\frac{(M \rho/2)^{2u} \cos\{\Delta u\}} 
{\Gamma[1+u] \Gamma[2+u]} \right] du   \nonumber \\
-2\left(\frac{7}{12}-\gamma\right) \sum_{k=1}^{\infty} (-1)^k 
\frac{(M \rho/2)^{2k} \cos\{k \Delta\} } 
{\Gamma[1+k] \Gamma[2+k] } .
\label{C}
\end{eqnarray} 
Some remarks concerning Eqs. 
(\ref{formula})
and (\ref{C}) are in order. The symbol $\sim$
indicates that the expansion
is to be understood only
in the assymptotic sense (see e.g. Ref. \cite{arf85}).
The $\#$ in the summation specifies that when
$\alpha$ is an integer, the summation
is done over every positive integer $k$, 
except
multiples of $\alpha$.
The ``$PV$" in front of the integration denotes its
Cauchy Principal Value.
These subtleties arise due to the fact that the function
$H_{\alpha}(x)$ defined in Eq. (\ref{F}) diverges for any $x$ which is a 
multiple of $\alpha$ . 
The Euler-Maclaurin formula used to derive
Eq. (\ref{formula}) has been carefully manipulated to
take into account these discontinuities. 

It is shown below that, for $M\rho\ll1$,
$C(M\rho/2,0)$ behaves as
\begin{eqnarray}
C(M\rho/2,0) = -\log(-\log(M \rho/2)^2).
\label{C1}
\end{eqnarray}

Writing Eq. (\ref{C}) for $\Delta = 0$ in terms of
\begin{eqnarray}
g(u) := \frac{\pi u}{\sin\{\pi u\}} \frac{1}{\Gamma[1+u] \Gamma[2+u]} \ \ \
\mbox{and} \ \ \ (M\rho/2)^{2u}, 
\end{eqnarray}
integrating by parts and rearranging the terms,
one has that 
\begin{eqnarray}
C(M\rho/2,0)= \gamma -g(0^+) \int_{0^+}^{1} \frac{1-(M\rho/2)^{2u}}{u} du  \
 + g(0^+) \int_{1}^{\infty} \frac{(M\rho/2)^{2u}}{u} du  \nonumber \\
 + \int_{0^+}^{\infty} \frac{g(u)-g(0^+)}{u} (M\rho/2)^{2u} du \
 + {\cal O}((M\rho/2)^2),
\label{C2}
\end{eqnarray}
where the ${\cal O}((M\rho/2)^2)$ is the leading behaviour of the
summation in Eq. (\ref{C}) for small values of $M\rho$.

The integrations in Eq. (\ref{C2}) behave, for small values of 
$M\rho$, as
\begin{eqnarray}
&&\int_{0^+}^{1} \frac{1-(M\rho/2)^{2u}}{u} du =  \log(-\log
(M\rho/2)^2)  +\gamma + {\cal O}(-\frac{(M\rho)^2}{\log M\rho }),
\nonumber  \\
&&\int_{1}^{\infty} \frac{(M\rho/2)^{2u}}{u} du  =  
{\cal O}(-\frac{(M\rho)^2}{\log M\rho }),  
\nonumber \\
&&\int_{0^+}^{\infty} \frac{g(u)-g(0^+)}{u}(M\rho/2)^{2u} du =  
{\cal O}(-\frac{1}{\log M\rho }). 
\label{eqs}
\end{eqnarray}

Thus, replacing $g(0^+)=1$ and Eqs. (\ref{eqs}) in Eq. (\ref{C2}),
one finally ends up with Eq. (\ref{C1}).

\end{document}